\documentclass[aps,twocolumn,showpacs,pre,preprintnumbers,groupedaddress]{revtex4}

\usepackage{graphicx}

\begin{document}

\title{Validity of the one-dimensional dissipative Boltzmann equation
for point particles up to the clustering regime}

\author{Jos{\'e} Miguel Pasini}
\email[]{jmp84@cornell.edu}
\altaffiliation{Present address: Department of Theoretical and Applied Mechanics, Cornell University,
Ithaca, NY 14853, USA}
\author{Patricio Cordero}
\homepage[]{http://www.cec.uchile.cl/cinetica/}
\affiliation{Departamento de F{\'\i}sica, Facultad de Ciencias
F{\'\i}sicas y Matem\'aticas, Universidad de Chile, Santiago, Chile}

\date{\today}

\begin{abstract}
We study stationary states of a one-dimensional gas of point-like
particles not subject to gravity between two walls at temperatures
$T_-$ and $T_+$, with $T_- < T_+$. Depending on the normalized
temperature difference $\Delta =
(T_+ - T_-)/(T_+ + T_-)$ the system may be completely
fluidized, or in a mixed state in which a cluster coexists
with the fluidized gas. We devise and explain in detail a
method for integrating the one-dimensional dissipative Boltzmann
equation in the test-particle limit for the stationary case.
We then apply this method to test the equation's validity up
to the clustering regime, by comparing with results from
microscopic Newtonian molecular dynamics. There is very good
agreement, with the one-particle phase space density
function presenting highly
non-Gaussian features, and a discontinuity that corresponds to
the test-particle limit. We conclude that Boltzmann's equation
is valid at least everywhere in the control parameter space where the
system has no cluster. The behavior of the system in its fluid
phase is dominated by characteristic lines which
resemble trajectories of particles subjected to a force which 
attracts them to a fixed point. If this point is
in the physical region a cluster forms, if not then the system
remains fluid.

\end{abstract}


\maketitle

\section{Introduction}

Granular systems have been the focus of much attention due both to the
theoretical challenges they present~\cite{kadanoff99} and to the
applications of industrial importance that stem from the rich phenomena
they exhibit (see Refs.~\cite{jaeger92,jaeger96,kadanoff99}
and references therein).
These systems are characterized by an energy loss in collisions, and
this loss is at the base of many interesting phenomena. Among these,
the clustering of particles has drawn much
attention~\cite{mcnamara92, mcnamara93, grossman96b, pasini01}.

In this paper we study in detail the mechanisms
that dominate the collective dynamics of a one-dimensional
system of point-like particles that interact via collisions
that conserve momentum but dissipate kinetic energy. A cluster
may or may not form. The system is confined in a box of unit
length and any particle that reaches a wall is expelled from
it with its velocity randomly chosen so that the
velocity distribution of ``outgoing'' particles is a
Maxwellian distribution with the ``temperature'' of
that wall. There are no external forces. In a previous
article~\cite{pasini01} we saw that there are two relevant
control parameters: the restitution coefficient which
characterizes the collisions and the normalized temperature difference
$\Delta$ between the walls, $\Delta=\frac{T_+-T_-}{T_++T_-}$. 
In the plane of these two parameters there is a {\em
transition line}: on one side the system is a granular fluid
that reaches a stationary regime while in the other side a
cluster is formed and, apparently, no stationary solution can
be reached, at least in the limit of infinitely many
particles. In Ref.~\cite{pasini01} we described what
happens, while in the present paper we disclose the underlying
mechanisms.

Two aspects have to be distinguished. On the one hand there is
the formal aspect: we show that Boltzmann's equation describes
closely what we get from molecular dynamic simulations
in the case of the pure fluid phase even quite close
to the transition line. On the other, there
is an intuitive picture---originated in our
detailed integration of Boltzmann's equation described in this
paper---that we explain in the following paragraphs.

To fix notation, if~$c_1$ and $c_2$ are the velocities of two
particles that are about to collide, their velocities after the
collision are given by
\[
c_1' = q c_1 + (1-q) c_2, \qquad c_2' = (1-q) c_1 + q c_2.
\]
Here $q = (1-r)/2$, where~$r$ is the usual restitution
coefficient. For the elastic case ($r=1$) the particles simply
exchange velocities. Since the grains are point-like, the
elastic case is then indistinguishable from a system in which
the particles do not interact. To make this more explicit,
the point-like character of the grains allows us to exchange
their identities after the collision, giving the collision rules
\[
c_2' = q c_1 + (1-q) c_2, \qquad c_1' = (1-q) c_1 + q c_2.
\]
Thus, when $q=0$ the velocities are unaffected, and when
$q$ is small the velocities are only barely changed. This 
leaves us with the picture of a system of weakly interacting
particles, whose relative velocity diminishes upon collisions.

The one dimensional granular system is being excited from the two
walls, generally at different temperatures, $T_-$ and $T_+$. 
Particles emerging from the walls act as
an outcoming ``wind'' pushing the particles away
from them. One could picture the effect of this wind as an
effective {\em repulsive force} which pushes the particles
away from the walls. If the temperature difference between the
two walls is large enough, the repulsive force associated to
the hotter wall prevails over the force associated to the colder
wall all across the system. Therefore in this case the overall
effect is a net force always pointing toward the colder wall,
much like gravity acts in a gas, always pointing to the base. If,
on the contrary, the temperature difference
is not large enough, there is a point in the
system where the two repulsive forces cancel
each other, producing an equilibrium point---a particle at rest
in this point would tend to remain at
rest---about which a cluster will grow. As the cluster
absorbs particles the density of the surrounding gas
decreases, and the equilibrium point may shift in time.

The article is organized as follows. In Sec.~\ref{sec:equation}
we state the kinetic equation in the limit which the authors of
Ref.~\cite{ramirez99a} call the \emph{hydrodynamic limit} of the
one-dimensional Boltzmann equation for point-like grains. This
limit is analogous to the Boltzmann-Grad limit of 
infinitely many  particles and finite mean free path. In
the present case, the role of the inverse of the mean free path
is played by the
factor~$qN$, where $N$~is the number of particles, and~$q$ is
the inelasticity factor mentioned above. This factor represents
how much is the velocity of a fast test-particle affected by
crossing the system, when the inelasticity~$q$ is very
small~\cite{du95, pasini01}. 
The boundary conditions and normalization used are also stated.

Section~\ref{sec:solution_algorithm}
focuses on the stationary state, and describes the algorithm devised
for the kinetic equation in this case.

Section~\ref{sec:results} contains results and conclusions. It compares 
the distribution functions in the
one-particle phase space~$f(x,c)$ (where~$x$ is the spatial coordinate,
and~$c$ is the velocity) with the same distribution measured from
molecular dynamic simulations. The numerical solution presents a
discontinuity that stems from the limit taken
in Sec.~\ref{sec:equation}, which allows the equation to be treated
as approximately linear. The measured distribution exhibits a
softened version of the discontinuity that steepens as we consider
systems with larger number of particles (thus approximating better the
limit of Sec.~\ref{sec:equation}). An intuitive  picture is
finally put forward.

\section{\label{sec:equation} Kinetic equation, boundary conditions, and normalization}

In the limit $N \rightarrow \infty$, but keeping $qN$ fixed, the
one-dimensional Boltzmann equation transforms into the test-particle
equation~\cite{mcnamara93,grossman96b,ramirez99a}:
\begin{equation}
\partial_t f + c \,\partial_x f = qN \partial_c (M f) ,
\label{eq:boltzmann_equation}
\end{equation}
where
\begin{equation}
M(x,c) = \int_{-\infty}^\infty f(x,c') (c - c') |c - c'| dc'.
\label{eq:M}
\end{equation}
Since the equation is nonlinear, we must define explicitly the
normalization used. In this case it is
\begin{equation}
\int_0^1 \int_{-\infty}^\infty  f(x,c) \, dc \, dx = 1.
\label{eq:normalization}
\end{equation}

The system is confined in a box of unit length, and the particles may
have any velocity: $(x,c) \in [0,1]\times (-\infty, \infty)$.
Any particle that reaches a wall is instantaneously expelled from
it---there is no adsorption---with its velocity randomly chosen so that the
velocity distribution of expelled particles is a Gaussian
distribution with the temperature of that wall. This
corresponds to choosing a wall kernel without memory and without a
delay time (see Ref.~\cite{cercignani90}, for example): 
\begin{equation}
f(0,c>0) \propto
e^{-c^2/2T_-} \qquad f(1,c<0) \propto e^{-c^2/2T_+} .
\label{eq:boundary_conditions}
\end{equation}
The temperatures at both walls are chosen so that the system temperature
for the perfectly elastic case is $T_0 = \sqrt{T_- T_+} = 1$. We will
always take $T_+ > T_-$.

The missing constants in Eq.~(\ref{eq:boundary_conditions}) are 
determined by Eq.~(\ref{eq:normalization}) and by imposing that there
is no flow across the walls:
\begin{equation}
\int_{-\infty}^\infty c \, f(x_{\mathrm{wall}}, c) \, dc= 0.
\label{eq:zero_flow_across_walls}
\end{equation}

\section{\label{sec:solution_algorithm}
Stationary state: solution algorithm}

In a stationary situation, Eq.~(\ref{eq:boltzmann_equation}) may be
rewritten as follows:
\begin{equation}
c \, \partial_x f - qN  M \partial_c f =  qN  f \, \partial_c M. 
\label{eq:stationary_case}
\end{equation}
The coefficient $-qN\,M $ multiplying $\partial_c f$ plays the
role of a force (per unit mass) and it is what we have called \emph{wind}, 
in the introduction. It is the effective total force on
a particle at $x$ with velocity $c$. When $f$ is reasonably
close to the true solution, $M$~will not depend on the detailed
form of the distribution. Thus, if we have a trial
distribution~$f_n$, we may consider $M$ and~$\partial M/\partial
c$ as given functions of $x$ and~$c$, and then we may 
solve Eq.~(\ref{eq:stationary_case}) for the
distribution~$f_{n+1}$. Seen in this light, when the trial
function is reasonably close to the solution,
Eq.~(\ref{eq:stationary_case}) is approximately a linear partial
differential equation that can be analyzed as a
hyperbolic equation, using the notion of characteristic
curves~\cite{couranthilbertvol2}. In the present case, the
characteristic curves satisfy
\begin{eqnarray} 
\frac{dx}{ds} &=& c \label{eq:dx_ds} \\
\frac{dc}{ds} &=& - qN  M(x,c) \label{eq:dc_ds} \\ 
\frac{df}{ds} &=& qN f \, \partial_c M(x,c) ,
\end{eqnarray}
where $s$ is a parameter.

In simple words, our integro-differential equation is treated as
if it were a (quasi-linear) partial differential equation and,
since real characteristics exist, it is possible to integrate
along these lines as if dealing with an ordinary differential
equation with a single independent variable~$s$.

Briefly put, given a distribution $f_n$ (which implies that
we have $M_n$ and $\partial_c M_n$), we calculate $f_{n+1}$ by
solving
\begin{equation}
c \, \partial_x f_{n+1} - qN M_n \, \partial_c f_{n+1} = qN
f_{n+1}  \, \partial_c M_n
\label{eq:iteratedequation}
\end{equation}
through numerical integration along the characteristics, following
Ref.~\cite{smith84}. After the integration we normalize $f_{n+1}$  to
one, and then use $f_{n+1}$ to calculate $f_{n+2}$. In this way we
eventually reach a fixed point.

\begin{figure}
\includegraphics[clip]{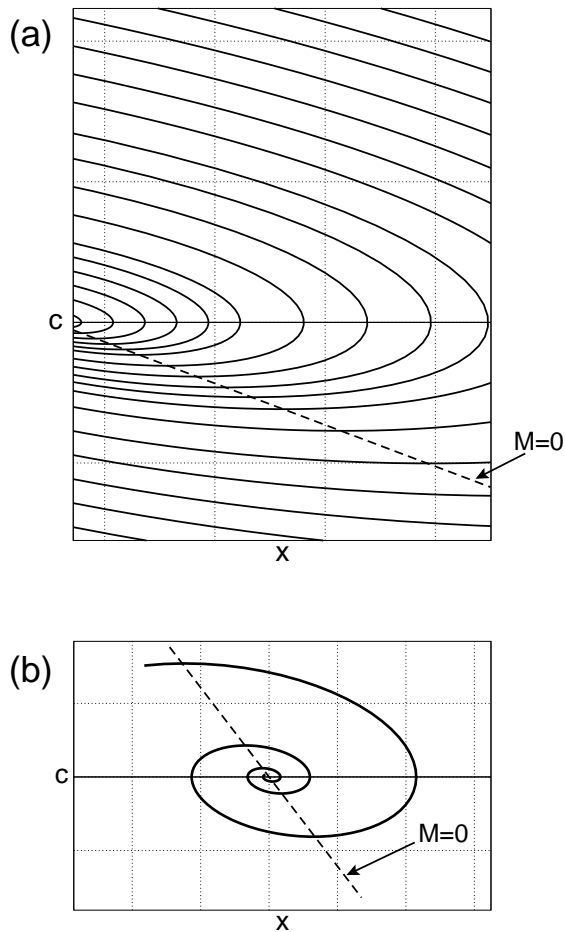}
\caption{Schematic representation of the form of the characteristic
curves in $(x,c)$-space when the right wall is hotter than the left
wall. Subfigure (a) shows the case where the curve $M(x,c)=0$ does not cross the $c=0$ line.
Subfigure (b) shows how a characteristic curve would wind around the point where the $M=0$ and
$c=0$ lines cross.
\label{fig:coiling_characteristics}}
\end{figure}

\begin{figure}[htb]
\includegraphics[clip]{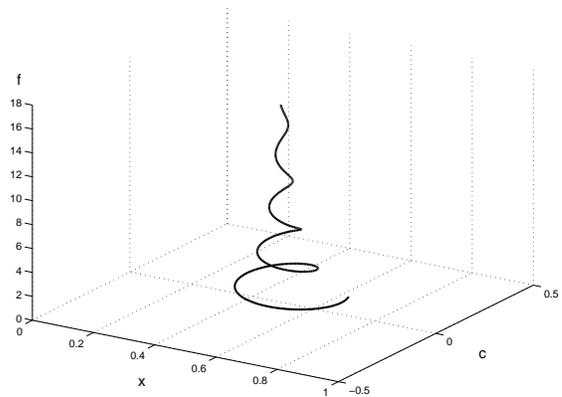}
\caption{Qualitative picture of the shape of a characteristic curve
in $(x,c,f)$-space  for
the case where the $M=0$ curve crosses the $c=0$ line.
\label{fig:winding_characteristic_x_c_f}}
\end{figure}

As will be described in detail in the following section,
there are three types of characteristics: (1) those that
originate at $x=0$ with a large positive velocity and end at
$x=1$; (2) those that also begin at $x=0$ (implying $c>0$)
but do not
reach $x=1$: they reach the $c=0$ axis, turn around,
and return to $x=0$; (3) those that start at $x=1$
(implying $c<0$) and end at $x=0$. The first
two types of characteristics are associated to the left boundary
condition, while the third type is associated to the right
boundary condition. The solution, therefore, may be
discontinuous along the separatrix of these last two types of
characteristics. Since our numerical algorithm integrates along
these characteristics, it never crosses the
discontinuity; every step deals with a smooth function.

As in Refs.~\cite{mcnamara93,grossman96b}  we may consider that
the projection of any characteristic line to the $(x,c)$ plane
corresponds to the phase-space trajectory of a \emph{test
particle} crossing the system. From Eq.~(\ref{eq:dc_ds}) we confirm
what has been said before, namely that $-qN M$ is the
acceleration of the particle. For large velocities $M \sim c\,|c|$,
hence if the particle's velocity is large it will be slowed
down. This is suggested in 
Fig.~\ref{fig:coiling_characteristics}(a), where the characteristics
far from $c=0$  approach that axis.

Due to the form of Eqs. (\ref{eq:dx_ds}) and~(\ref{eq:dc_ds}),
the intersection of the curves $c=0$ and $M(x,c)=0$ is
interesting~\footnote{Since $M \sim c\,|c|$ asymptotically, we can be sure
that there is at least one $M=0$ curve, and that if there is only
one curve, then ``above'' it $M>0$, and ``below'' it $M<0$.
Therefore in this case $\partial_c M > 0$ at the intersection of
$M=0$ and $c=0$.}.
At $c=0$ the characteristic curves in the $(x,c)$ plane
are vertical, and at $M=0$ they are horizontal.
Figure~\ref{fig:coiling_characteristics}(b) is a sketch of what
would happen to the characteristic curves around the intersection
point~(which we will call $G$ henceforward): they would wind
around it, never reaching it. Meanwhile, since
$\partial_c M > 0$ in that vicinity, we have that $f$ is increasing along
the curve. In other words, in $(x,c,f)$-space the characteristic curve
is practically vertical, with $f$ increasing sharply along it. This case,
depicted qualitatively in Fig.~\ref{fig:winding_characteristic_x_c_f},
corresponds to the presence of a cluster at the intersection point~$G$.

Hence, for the system to be able to reach a stationary fluidized state, the
curve $M=0$ must never cross the $c$ axis. In
Fig.~\ref{fig:coiling_characteristics}(a) we have drawn this curve in the
lower part of the plane, which corresponds to the case when the
left wall is the colder one ($T_-<T_+$) and the temperature difference is
large enough that no cluster can form.

Further details of the algorithm are relayed to the appendix. 

\section{\label{sec:results}Results and conclusions}

\begin{figure}[htb]
\includegraphics[clip]{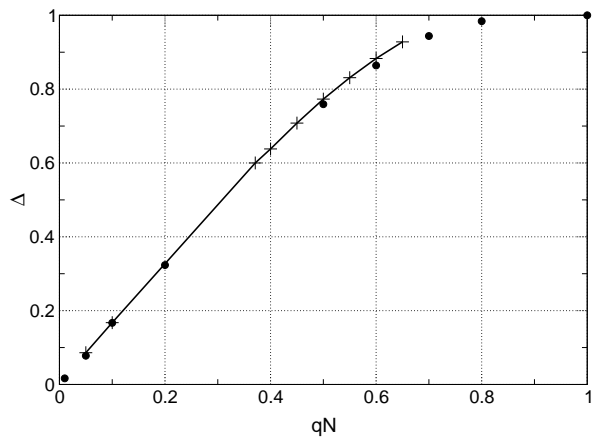}
\caption{Comparison between the threshold for cluster formation in
molecular dynamics and the loss of convergence in the algorithm. The
plus signs show the lowest values of~$\Delta$ for each~$qN$ before the
algorithm becomes unstable.  The circles show the lowest values
of~$\Delta$ for each~$qN$ before a cluster is detected in a molecular
dynamics simulation of $N=1000$ particles. The plus signs are joined by
lines to guide the eye. \label{fig:threshold}}
\end{figure}

\begin{figure}[htb]
\includegraphics[clip]{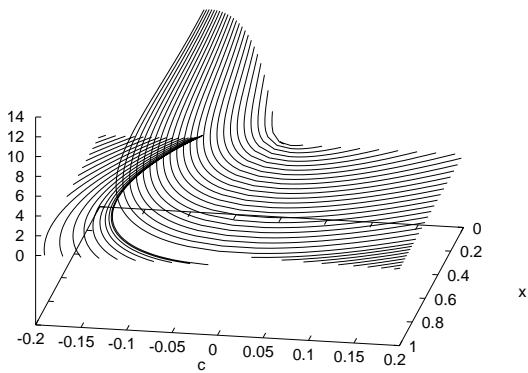}
\caption{Distribution function $f(x,c)$ for the case $qN=0.35$ and
$\Delta = 0.6$. Only one every four curves is shown to unclutter the
picture.
\label{fig:sparse_f_qN035delta06}}
\end{figure}

\begin{figure}[htb]
\includegraphics[clip]{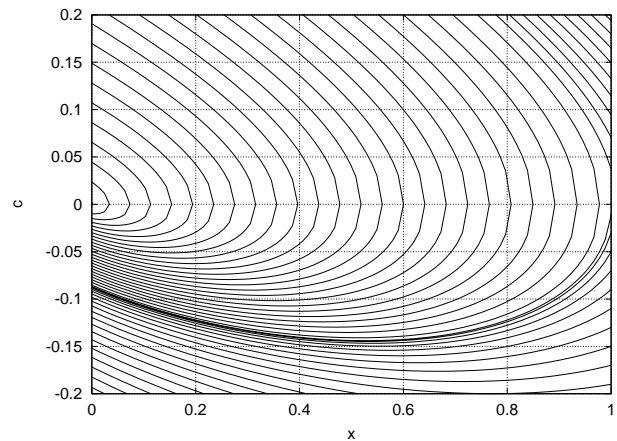}
\caption{Projection of the characteristic curves into ($x$,$c$)-space
for the same case as Fig.~\ref{fig:sparse_f_qN035delta06}. Only one
every four curves is shown to unclutter the picture.
\label{fig:sparse_characs}}
\end{figure}

Figure~\ref{fig:threshold} shows the threshold for cluster formation in
molecular dynamics and the loss of convergence in the algorithm.
We can say that they coincide, namely, the integration of
Boltzmann's equation with the present algorithm converges almost 
to the transition line beyond which a cluster begins to form.

Figure \ref{fig:sparse_f_qN035delta06} shows the characteristic
lines in $(x,c,f)$-space when $qN=0.35$ and $\Delta=0.6$. Although it 
corresponds to a completely fluidized case it is not to far from the
transition line, as can be seen in Fig.~\ref{fig:threshold}.

\begin{figure}[htb]
\includegraphics[clip]{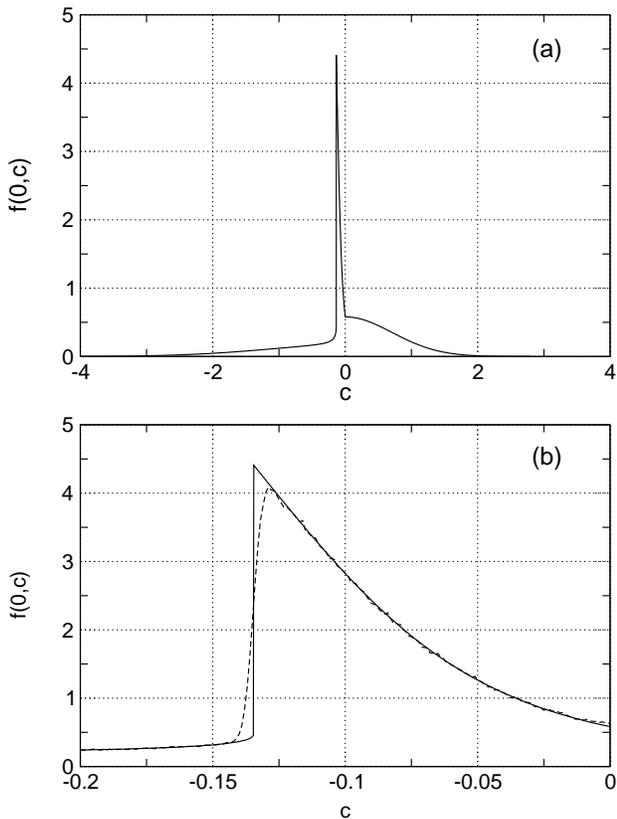}
\caption{Distribution function at the cold wall for the case $qN=0.1$ and $\Delta = 0.6$.
The solid curve results from applying the algorithm described in the
text, while the dashed curve was measured from a molecular dynamics
simulation with $N=1000$ particles. Subfigures~(a) and~(b) show the
same data, but on a different scale. Subfigure~(a) emphasizes how the
distribution is essentially Gaussian, save for the sharp peak for slow
particles. Subfigure~(b) shows a detail of this peak, showing how the
discontinuity is smoothed in the simulation.
\label{fig:distribution_and_peak_detail}}
\end{figure}

Figure~\ref{fig:sparse_characs} shows the projection of
the characteristic curves into the $(x,c)$ plane.
The asymmetric aspect of the family of curves is
due to the effective force which pushes
particles toward the colder wall almost as if there was a
space-dependent gravity-like force. The figure only shows the characteristic 
lines for small
values of the velocity since further away the distribution is a
bimodal Gaussian. One interesting feature of this figure is the density
of lines near $x=0$ for small negative velocity. It corresponds
to a remarkable peak of the velocity distribution for velocities
much smaller than the thermal velocity. In the present case the
thermal velocity for the particles approaching the wall is about
$\sqrt{2}$, an order of magnitude larger than the width of the
peak.

%% THE SQRT(2) IS CORRECT

To check how the distribution obtained from our
algorithm compares with the distribution stemming from molecular dynamic
simulations we show the distribution at the colder wall
$f(0,c)$, where the discontinuity is more notorious.  In
Fig.~\ref{fig:distribution_and_peak_detail} we compare
with molecular dynamic results for
$N=1000$ particles, $qN=0.1$, and $\Delta=0.6$, namely quite far
from the clustering regime. 
Figure~\ref{fig:distribution_and_peak_detail}(a)
shows the distribution for a wide range of velocities (at this
scale the calculated and the simulated solutions cannot be
resolved by the naked eye). It can be
checked that the distribution behaves as two Maxwellians, one for
$c>0$ and a different one for $c<0$ (with $|c|$ sufficiently large),
and has a remarkable peak for small negative velocities. 
Even though the system is
far from the clustering regime, that peak is a reminder that a
clustering singularity exists.

\begin{figure}[htb]
\includegraphics[clip]{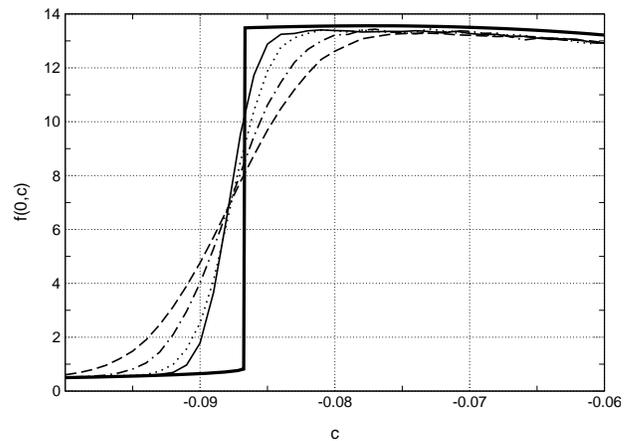}
\caption{Detail of the discontinuity in the distribution function at
the left wall for $qN=0.35$ and $\Delta = 0.6$. The thick solid line 
results of applying the algorithm described in the text. The other
curves stem from molecular dynamics for 1000 (dashed), 2000
(dash-dotted), 5000 (dotted), and 10000 (thin solid line) particles.
\label{fig:size_effects}}
\end{figure}

Figure~\ref{fig:distribution_and_peak_detail}(b) shows in detail
the shape of the discontinuous behavior at the peak. The discontinuity
that the analytic analysis predicts is softened in the
simulations. This difference is due to size effects. In fact,
as seen in Fig.~\ref{fig:size_effects} (which corresponds to a
system on the verge of clusterization), simulations of systems with
an increasing number of grains exhibit increasingly steeper
behavior at the predicted discontinuity, approaching the
result that Boltzmann's equation implies and our algorithm yields.

The final picture that emerges from all that has been said is,
first, that Boltzmann's equation for the quasi-elastic system 
is valid essentially in all
points of the $(q,\Delta)$ plane where the system has no
cluster. The behavior of the system in its fluid phase is
dominated by characteristic lines, trajectories of a {\em test
particle} subjected to a force which attracts it to a 
point $G$ where the $M=0$ line crosses the $c$-axis. In the fluid
phase such point $G$ is beyond the physical box, particles hit
the colder wall, forget their past and reenter the system.

We have not solved the time-dependent case when $G$ is inside
the box. In such case many trajectories in phase space will wind
around $G$ (as shown in Figs.~\ref{fig:coiling_characteristics}(b)
and~\ref{fig:winding_characteristic_x_c_f}),
and the density at that
point will tend to diverge,
thus forming a cluster. 

\begin{acknowledgments}
This work has been partially funded by Fundaci{\'o}n Andes, by
\emph{Fondecyt} grants No.~2990108 and  No.~1000884, and by 
\emph{Fondap} grant 11980002.
\end{acknowledgments}

\appendix*

\section{Algorithm in detail}

In order to commence the integration, we need some initial
\emph{ansatz} $f_0(x,c)$. This may be,
for example, the solution for the elastic case:
\[
f(x,c) =  A\left\{\theta(c) e^{-c^2/2T_-} + B[1 - \theta(c)] e^{-c^2/2T_+}\right\},
\]
where $\theta(c)$ is the Heaviside step function, $B$ is a constant
chosen so as to have zero flux at the walls, and $A$ is a normalization constant. 
In practice, if it is available, it is convenient
to choose as $f_0$ a previous solution for a case similar to the one being
studied. This not only speeds up the convergence, but also
may help avoid the spurious appearance of $G$ points at
intermediate iterations.

The next step is to see whether the $M=0$ curve crosses the $c=0$ line.
If it does, the algorithm breaks down, and we must start from a better
\emph{ansatz}. If it lies wholly on $c < 0$, the characteristic curves
will behave qualitatively as in Fig.~\ref{fig:coiling_characteristics}(a).
In this case we start by integrating the characteristics that begin
at $(x=0, c>0)$. If the curve lies wholly on $c > 0$ the situation is
inverted, and we must start from $(x=1, c<0)$. In what follows we
will assume that the situation is as in Fig.~\ref{fig:coiling_characteristics}(a).
This is always the case when we are sufficiently close to the solution.

The next step is finding the dividing characteristic. This is done directly by
the bisection method, choosing different values of $c_D$ so that the
initial condition for the characteristic is
\[
x(s=0) = 0, \qquad c(0) = c_D, \qquad f_1(0) = A_- e^{-c_D^2/2T_-}.
\]
The value of $A_-$ is chosen so that the net flux at the wall is zero:
\[
\int_{-\infty}^0 c f_0(0,c)dc + A_- \int_0^\infty c \,e^{-c^2/2T_-}dc = 0.
\]
Integrating along this characteristic in very small steps we find $c_D$
such that the corresponding curve has a turning point at $(x,c)
= (1- \epsilon, 0)$, with $\epsilon$ less than a reasonably small number.

With this value of $c_D$ we divide the interval $0 < c < c_D$,
choosing many values $c_i$ (with $0 < c_1 < c_2 < \ldots < c_D$) 
as starting points for characteristic curves. These will all be curves which will turn
around and return to $x=0$. The crossing points will define a
natural discretization of the $x$~axis which will be used to
tabulate the values of $f_1$. In this way we always sample the turning
point.

Now we start the integration: with \emph{increasing} $i$, we integrate the
characteristic that starts with $c_i$, tabulating the values of $c(s)$
and $f(s)$ for values of $x$ corresponding to the turning points of
the previously integrated characteristics. This step is of paramount
computational importance: at the next iteration we will need
to calculate the integrals $M(x,c)$ and $\partial_c M(x,c)$.
Since these are integrals in $c$ keeping $x$ fixed, it is
important to have values of $f(x,c)$ tabulated at the same values
of~$x$. This precaution allows us to evaluate $M$ and $\partial_c M$
in a fast and straightforward manner.

Having integrated along all the characteristics with $c_i < c_D$, we
now integrate characteristics that start from $(x=0,c>c_D)$. These
are curves that start at the left wall and reach the right wall, that
is, they do not have a turning point.
The completion of this step implies that we have $f_1$ in the whole
half-space $c>0$ (and also in part of $c<0$). 

At this point we can start integrating the characteristic curves
that start at the right wall, choosing values of $c_0 < 0$:
\[
x(s=0) = 1, \qquad c(0) = c_0, \qquad f_1(0) = A_+ e^{-c_0^2/2T_+}.
\]
The value of $A_+$ is chosen so that the net flux at the wall is zero:
\[
A_+ \int_{-\infty}^0 c \, e^{-c^2/2T_+} dc + \int_0^\infty c f_1(1,c) dc = 0.
\]

After completing the integration, we normalize $f_1$, and take the magnitude
of this adjustment as a measure of how far we are from reaching a fixed point.

This procedure is repeated until a reasonable convergence is reached.
Typically ten to fifteen iterations are necessary to achieve a good solution.

\end{document}